\documentclass{./aa}     
\usepackage{graphics}

\begin{document}
                   
\title{
 Sidereal time analysis 
as a tool for the study of the 
space distribution of sources of gravitational waves
}
\titlerunning{Sidereal time analysis}
\author{ G. Paturel\inst{1} and Yu. V. Baryshev\inst{2,}\inst{3}}

\offprints{G. Paturel}

   \institute{
 CRAL-Observatoire de Lyon,
 F69561 Saint-Genis Laval Cedex, France \\
\and
 Astronomical Institute of the Saint-Petersburg University,
 198504, St.-Petersburg, Russia\\
\and
  Isaac Newton Institute of Chile, St.-Petersburg Branch, Russia\\ 
}

   \date{Received 15 July 2002; Accepted 15 October 2002 }

\abstract{
Gravitational wave (GW) detectors operating on a long time range
can be used for the 
study of space distribution of sources of GW bursts
or to put strong upper limits on the GW signal of a wide class 
of source candidates.
For this purpose we propose here a sidereal time analysis to
analyze the output signal of GW detectors.
Using the characteristics of some existing  detectors, 
we demonstrate the  capability of the sidereal time analysis to give
a clear signature of different localizations of GW sources:
the Galactic Center, the Galactic Plane, the Supergalactic
plane, the Great Attractor.
On the contrary, a homogeneous 3D-distribution of GW sources
gives a signal without features.
In this paper we consider tensor gravitational waves with randomly oriented
polarization. We consider GW detectors at fixed positions on the
Earth, and a fixed orientation of the antenna.}
\keywords{
gravitational -
waves --
methods: data analysis --
methods: statistical
               }

   \maketitle
\section{Introduction}
In this paper, we base our study of detection of gravitational waves (GW)
on a phenomenological approach as formulated by Finn (2001). 
We explore the reality that could emerge directly from 
observation, without considering specific models of GW emission.
As has happened frequently in the history of astronomy when a new observational
window opens, GW astronomy may lead to a revision of theoretical models and 
to the development of new ideas. 
Even upper limits on the energy of GW events are very informative because
they put restrictions on the predicted characteristics of GW sources and then
on the physics of GW themselves.
Energetic GW bursts often arise in regions where the gravity effects
are strong (supernova explosion, neutron star merging...).  They thus carry 
information about fundamental gravity physics in an unexplored  regime of
strong field.

Our phenomenological approach is also justified by the fact
that standard calculations of GW signals from supernova explosions and merging of
relativistic compact massive objects essentially depend on the details
of the last relativistic stages of the gravitational collapse that are
still poorly known (Thorne 1987,1997; Paczynsky 1999; Finn 2001). 

Several GW detectors of the third generation are 
operating with long duty times.
Five resonant GW detectors of the International Gravitational
Event Collaboration (IGEC) already cover 1332 days of observation (Astone 2001),
i.e. about 90\% of the total time between 1997-2000.
In 2002, the laser interferometric detectors TAMA300, GEO600
and LIGO started to operate, though with shorter duty cycles.

Usual method of data analysis for
GW detectors consists of searching for coincidences of events between 
two or more detectors (Prodi et al. 2000).
Here we consider another method $-$ sidereal time analysis (STA) $-$.
that consists of recording all GW events as a function of
sidereal time.
This possibility was mentioned by Weber (1969) in his pioneering work.
The Earth's rotation, measured with respect to stars, gives a natural 
unit of time for ground-based GW detectors.
The advantage of the STA method is that the noise events do not
correlate with sidereal time, while true GW signals, coming
from a certain distribution of GW sources, is tightly related
to definite sidereal time. This happens because, for a fixed
distribution of GW sources, a given antenna pattern produces 
a fixed amplitude shape along the sidereal time.
Hence, for a given detection threshold,  GW events
will be detected only for specific sidereal time intervals.

Here, considering tensor waves first, we demonstrate the capability of
the STA method to give a clear signature of some realistic distributions 
of GW sources. 
For this phenomenological approach
we must choose a very general form of the GW burst amplitude and then
a distribution of GW sources.
In section 2 we explain how the amplitude of individual sources
is chosen and how the considered distributions of sources
are built.
In section 3 we give, for each distribution, the relative rate of 
GW events along sidereal time, by considering 
a Weber-type bar-detector (EXPLORER) and 
an interferometric detector (VIRGO).
In section 4 we analyze the results to show how to use our results for
a practical comparison with real observations. 

Our main target is to prepare the tools for the concrete analysis of
forthcoming GW signals. Our FORTRAN program is available on request.

\section{Description of normalized sources}

\subsection{Amplitude of an individual GW burst emitter}

Let us recall briefly how the count of events is made. For more details
see Baryshev and Paturel (hereafter, BP2001).

Depending on the energy transformed into GW by a source,
one will observe a change in the reference length $l$. 
This is measured by the observed amplitude  $h_{obs}=\Delta{l}/l$. 
This signal will be counted as a detected event only if $h_{obs}$ is
larger  than the limiting amplitude $h_{lim}$ that defines the 
sensitivity of the detector.

Many of the expected signals may be presented
as the "unit pulse of GW burst" introduced by
Amaldi \& Pizzella(1979). 
It is a GW burst in a sinusoidal wave with amplitude $h_0$, frequency $\nu_0$
and duration $\tau_g$. For tensor GW, the observed amplitude $h_{obs}$ 
at a distance $r$, produced by energy $E_{gw}$, is (see e.g. Pizzella 1989):

\begin{equation}
h_{obs} =1.4 \times 10^{-20} (\frac{1 Mpc}{r})(\frac{10^3 Hz}{\nu_0})
(\frac{E_{gw}}{1 M_{\odot}c^2})^{1/2}(\frac{1 s}{\tau_g})^{1/2}
\label{ho}
\end{equation}
Hence, a GW source at a fixed distance $r$, to a first approximation,
is characterized by three main observable parameters
$E_{gw}$, $\nu_0$ and $\tau_g$, for a wide class of radiation processes.

In the present paper, we consider the normalized amplitude that may be
written as (r in Mpc): 

\begin{equation}
\frac{h_{obs}}{h_{\star}} \propto \frac{|G(\alpha, \delta)|}{r}
\label{hobs}
\end{equation}

The term $G(\alpha, \delta)$ is a geometrical factor that defines the
position of the source, with respect to the 
detector. The expression of this term is given for bar and interferometric 
detectors in BP2001.
The quantity $h_{\star}$ defines the type of source.
For galactic weak sources we will adopt $h_{\star}=10^{-20}$ or 
$h_{\star}=10^{-23}$.
For powerful extragalactic sources of GW, like supernovae, we will adopt 
$h_{\star}=10^{-17}$ or $h_{\star}=10^{-19}$. 
Obviously, our calculations may be rescaled by using any
other combinations of main GW parameters.

Because the polarization angle cannot be predicted,
it has been taken at random.  The net effect of this unknown polarization
is that the number of events, at a certain level above the noise, 
is simply reduced (for more detail see BP2001). This does 
not change the shape of the count rate with sidereal time. Note that the
polarization makes it that one cannot infer the true amplitude of a GW event
from the observed amplitude. 
Nevertheless, a large amplitude means both a large true
amplitude and a favorable polarization.

\subsection{Distribution of sources}

We now define a group of individual GW sources. 
The sources inside such a group have a well-defined distribution.
For instance, a group can be the center of our Galaxy. In this case each individual 
source is a galactic emitter (like neutron star mergers or unknown sources of 
GW pulses) distributed in a sphere representing the region around the center of our Galaxy. 
Another example of a group can be a supercluster of galaxies. In this
case each individual source is a galaxy. 
The group can be also a flat distribution of emitters with a given thickness
and a given size.  An example is de Vaucouleurs's Local Super Cluster
(de Vaucouleurs, 1956, 1975).

Each group of emitters is characterized by:
\begin{itemize}
\item Number of individual emitters. Generally, this number is arbitrary
but, when one considers several groups, their relative populations
must be fixed.
\item The rate of GW burst emission, e.g. in solar masses per unit of time.
Because we calculate the relative number of detected events (in percent) 
for an arbitrary
number of sources, the rate of GW emission for each individual emitter 
may be set to one. This assumes that each individual
emitter has the same probability of producing a GW event.
\item The coordinates of each individual emitter. We adopted equatorial coordinates.
So, one can use an actual catalog of emitters (like a catalog of real
galaxies) or a simulated catalog, using coordinates obtained
from an artificial distribution of emitters. 
\item Distance of each emitter. 
For a simulated sample the distance is given by the considered
distribution.
For an observed catalog of galaxies, 
the distance can be deduced from the radial velocity using a given Hubble
constant. 
\end{itemize}

There are several natural space distributions of GW sources
that may be directly tested by the STA method.
Here we consider the following models of GW source distributions that
will be tested in the next section:
\begin{enumerate}

\item Homogeneous 3D-distribution. We consider a homogeneous distribution of 10000 emitters
in a sphere of radius $R=10 kpc$ centered on the detector. 
Each emitter is considered as a galactic object with a
reference amplitude $h_{\star}=10^{-20}$ or  $h_{\star}=10^{-22}$ for EXPLORER or VIRGO,
respectively. This distribution will be used to show that,
on average, there is a constant rate of GW events along sidereal time.
Obviously, the result is the same when a homogeneous 3D-distribution of
extragalactic objects is used (we made this verification).

\item Galactic center. The galactic center is represented by one group of 10000 emitters
located at the 1950 equatorial coordinates: $B174224.0-285500$. The number of sources
is assumed to decrease following a Gaussian distribution with standard deviation (radius)
 of $\sigma=0.001 Mpc$. 
The adopted distance between the Sun and the galactic center is $0.010 Mpc$.
The kind of emitter is
not explicitly identified, but one can imagine that the galactic center could be
the source of many strong GW events.
A reference amplitude $h_{\star}=10^{-20}$ is used for the kind of emitters detectable
with EXPLORER and $h_{\star}=10^{-23}$ for emitters detectable
with VIRGO.

\item The galactic plane. The distribution will be centered on the galactic center
$B174224.0-285500$ at a distance of $0.010 Mpc$ from the Sun. The radius of the
distribution (radius of our Galaxy) is taken as $0.0125 Mpc$. The adopted thickness
of the disk is $2kpc$. We consider 10000 emitters with a reference amplitude of
$h_{\star}=10^{-20}$ or $h_{\star}=10^{-23}$ for EXPLORER and VIRGO respectively. 
Each emitter will be distributed homogeneously in this
thin disk.
Note that, because our Sun is not at the center of this distribution, the
distribution along the galactic longitude is not uniform. Again, for this distribution, 
we do not use {\it a priori} information on the types of GW sources.

\item The supergalactic plane. The dominant structure around the Local Group is
de Vaucouleurs Local Super Cluster (LSC). The Virgo galaxy cluster is the dominant
structure of the LSC. We will consider a homogeneous distribution
of 10000 emitters in a disk of radius $25 Mpc$ and height $5 Mpc$, centered on the
Virgo cluster itself. 
The north pole of the supergalactic plane is at $l=47.37$ degrees, $b=+6.32$ degrees.
The reference
amplitude of each emitter will be  $h_{\star}=10^{-17}$ or $h_{\star}=10^{-19}$ for
EXPLORER and VIRGO respectively.
According to  recent work by  Norris (2002) the faint, slowly-cooling Gamma Ray 
Bursts could be localized in the supergalactic plane. Although the effect is not 
yet confirmed it is interesting to predict what would be the rate of sources 
localized in the supergalactic plane.

\item Virgo and the possible Great Attractor. This distribution is made of 1000 emitters
(galaxies) in the Virgo cluster, centered  at the 1950-equatorial coordinates
$B122800.0+124000$. We assumed a distance of $17Mpc$ and a radius of $1.7Mpc$.
We expect that Virgo will produce the most easily detectable events (this is the
reason why the Franco-Italian detector is called VIRGO). However, it has been
suspected that a very huge structure, the Great-Attractor (Dressler et al., 1987; hereafter, GA), 
could be present at
three times the distance of Virgo, but hidden by our Galaxy. From the acceleration
induced on Virgo, one can assume that the GA has 27 more galaxies than Virgo. We thus
added to our synthetic sample 27000 galaxies, centered at the 1950-equatorial coordinates
$B150000.0-600000$. We assumed a distance of $51 Mpc$ and a radius of $5.1 Mpc$
For each emitter of both Virgo and the GA we assumed that the reference amplitude
is $h_{\star}=10^{-17}$ or $h_{\star}=10^{-19}$ for EXPLORER and VIRGO respectively. 
With a high sensitivity detector, we expect that the GA
could become the most efficient source of GW events. This could give a clue for
the detection of this putative dense region.

\end{enumerate}

\subsection{Practical method for constructing the samples}
\begin{figure}
\resizebox{\hsize}{!}{\includegraphics{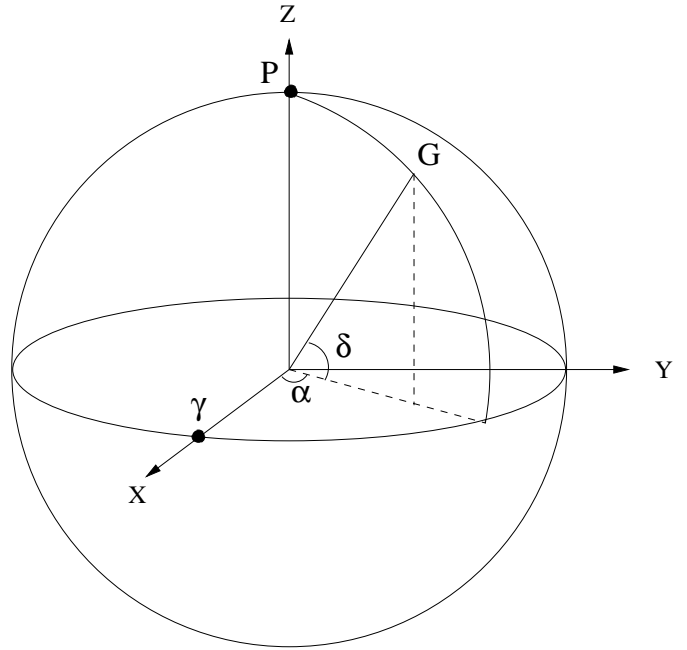}}
\caption{Definition of the coordinates. The random $X,Y,Z$ coordinates define
the position of an individual emitter $G$ (e.g., galaxy). The $X$-axis gives 
the origin of Right Ascension $\alpha$ ($\gamma$ direction). The $XY$-plane defines the
equatorial plane used to define the declination $\delta$. 
Each random emitter $G$ is accepted (or rejected) for obeying
the searched distribution.
}
\label{xyz}
\end{figure}

The Cartesian $X,Y,Z$ coordinates of each emitter are chosen at random, for a uniform
distribution, or following a Gaussian distribution around the center, for a discrete group 
like a cluster. These coordinates are chosen according to Figure \ref{xyz}.
The $X$-axis gives the direction of the origin of Right Ascension ($\gamma$ point).
The North equatorial pole is defined by the $Z$-axis. The $X,Y,Z$ coordinates are then
transformed into equatorial coordinates and distance.
Then the coordinates are transformed into galactic or supergalactic coordinates,
depending on the distribution we are constructing.
An individual emitter is then accepted, or not, depending on its position with
respect to the distribution. This operation is repeated until the number of
emitters has reached the desired value.

We considered the following detectors:\\
\vspace{0.3cm}
\begin{tabular}{llrl}
Detector        & Latitude $L$&  Azimuth $\Phi_o$  & Type \\
\hline
EXPLORER        &$46.5\deg $&$ -39\deg $& bar     \\
VIRGO           &$43.6\deg $&$ -19\deg $& interf. \\
\hline
\end{tabular}
\label{detector}
\vspace{0.3cm}

The considered sensitivity range for EXPLORER is $10^{-18}$ to $10^{-20}$
and for VIRGO $10^{-20}$ to $10^{-22}$
The main useful characteristics are: the {\it latitude} on the Earth
(let us recall that the longitude on the Earth does not intervene because 
the count rate of GW events is registered as a function of sidereal time), the {\it azimuth}
of the main axis (the main axis of the bar or the x-arm for an interferometric
detector) and the {\it type of detector}.

\section{Results}

We calculated the GW normalized count rates as a function of sidereal time
for the two detectors listed above.
For each of the studied distributions we give a figure with three panels:
\begin{enumerate}
\item Panel 1 (top): A Flamsteed equal-area projection that shows the distribution in supergalactic coordinates.
\item Panel 2 (middle): The shape of the normalized count rate predicted for the EXPLORER bar 
detector along the sidereal time axis. 
We consider five sensitivity levels: 
$10^{-18.0}$, $10^{-18.5}$, $10^{-19.0}$, $10^{-19.5}$ and $10^{-20.0}$. The corresponding
curves appear from the bottom to the top; the higher
the sensitivity the higher the rate of events.
\item Panel 3 (bottom): The shape of the normalized count rate predicted for the VIRGO 
interferometric detector along the sidereal time axis.
We consider five sensitivity levels: 
$10^{-20.0}$, $10^{-20.5}$, $10^{-21.0}$, $10^{-21.5}$ and $10^{-22.0}$. The corresponding
curves can be identified as said for panel 2.
\end{enumerate}

\begin{figure}
\resizebox{\hsize}{!}{\includegraphics{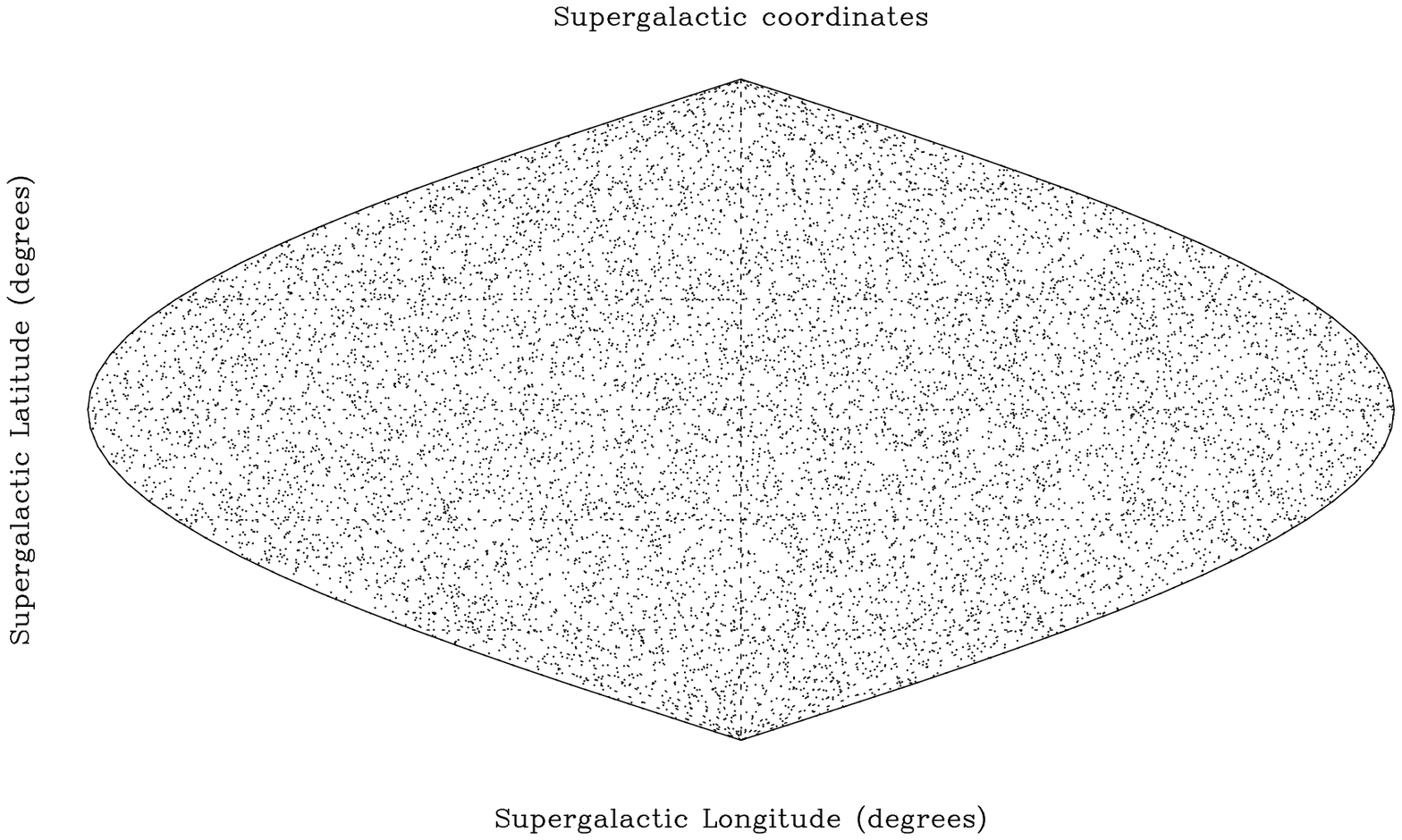}}
\resizebox{\hsize}{!}{\includegraphics{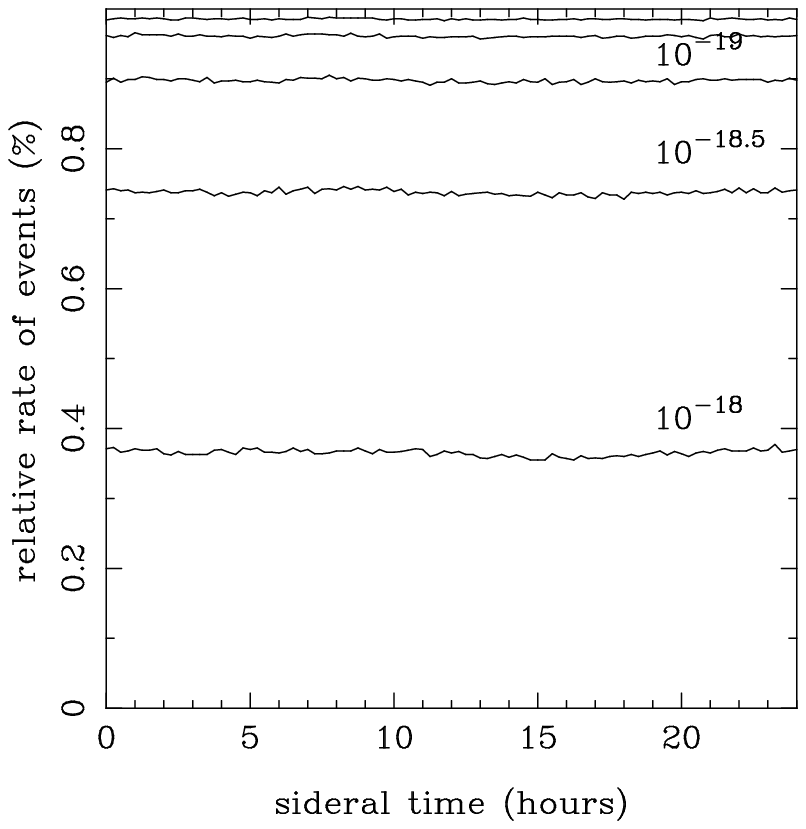}}
\resizebox{\hsize}{!}{\includegraphics{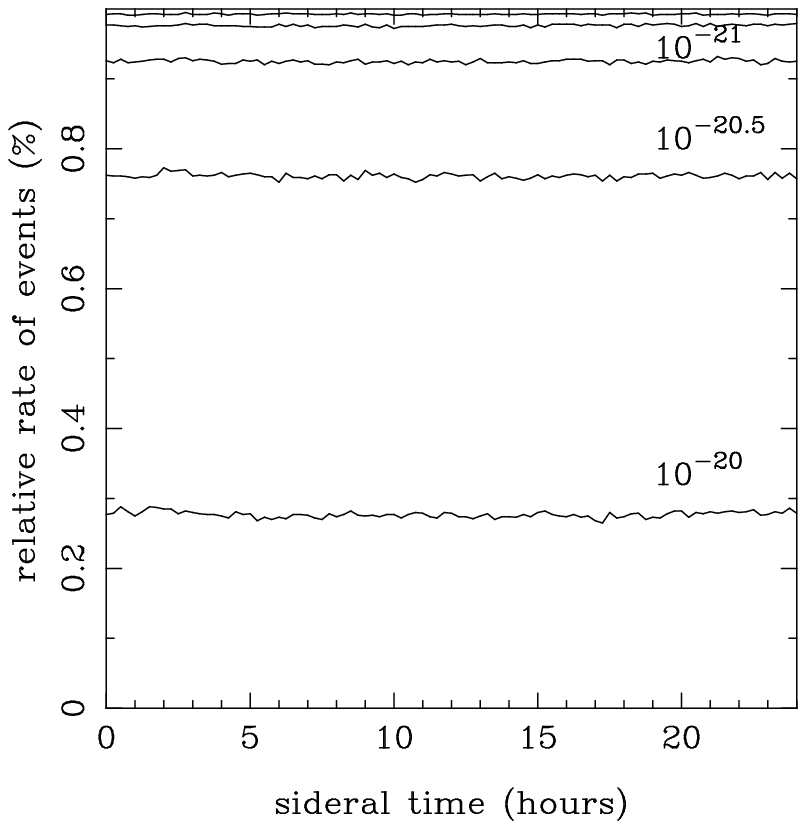}}
\caption{3D homogeneous distribution of sources within 10 kpc. 
See section 3 for explanations.}
\label{fi1}
\end{figure}

\begin{figure}
\resizebox{\hsize}{!}{\includegraphics{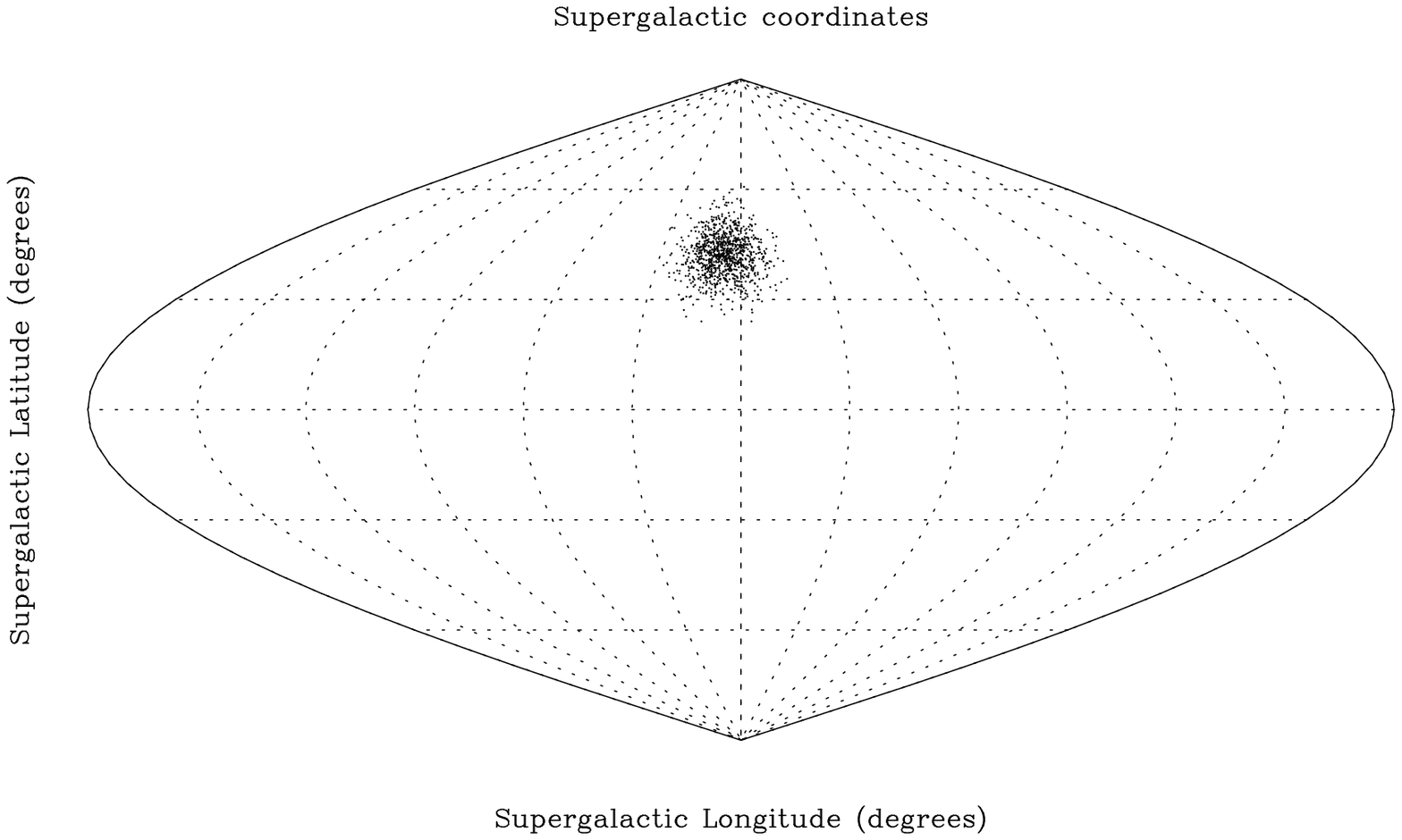}}
\resizebox{\hsize}{!}{\includegraphics{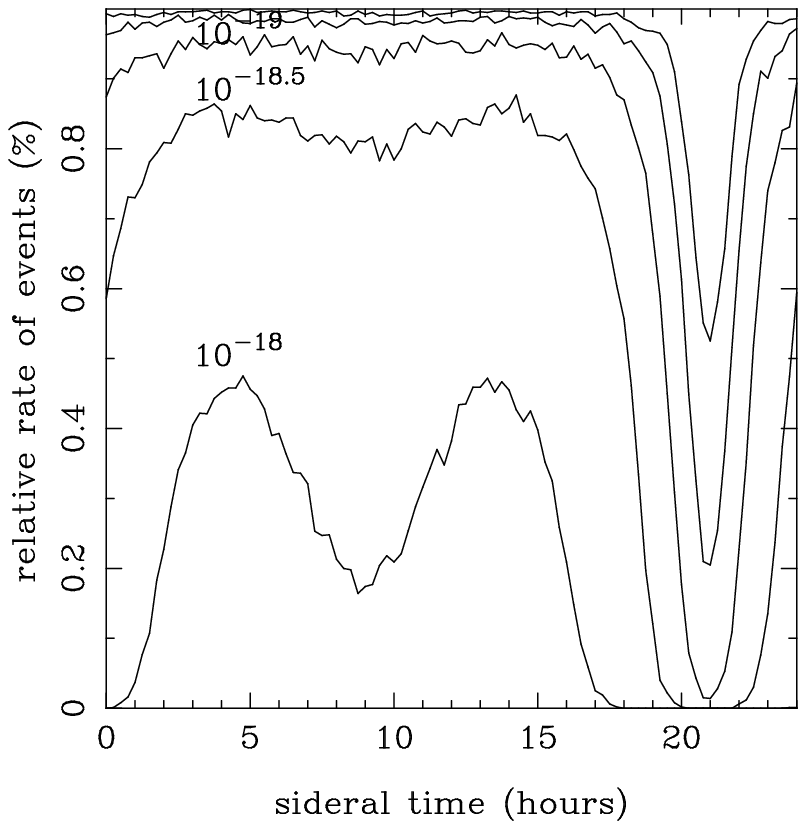}}
\resizebox{\hsize}{!}{\includegraphics{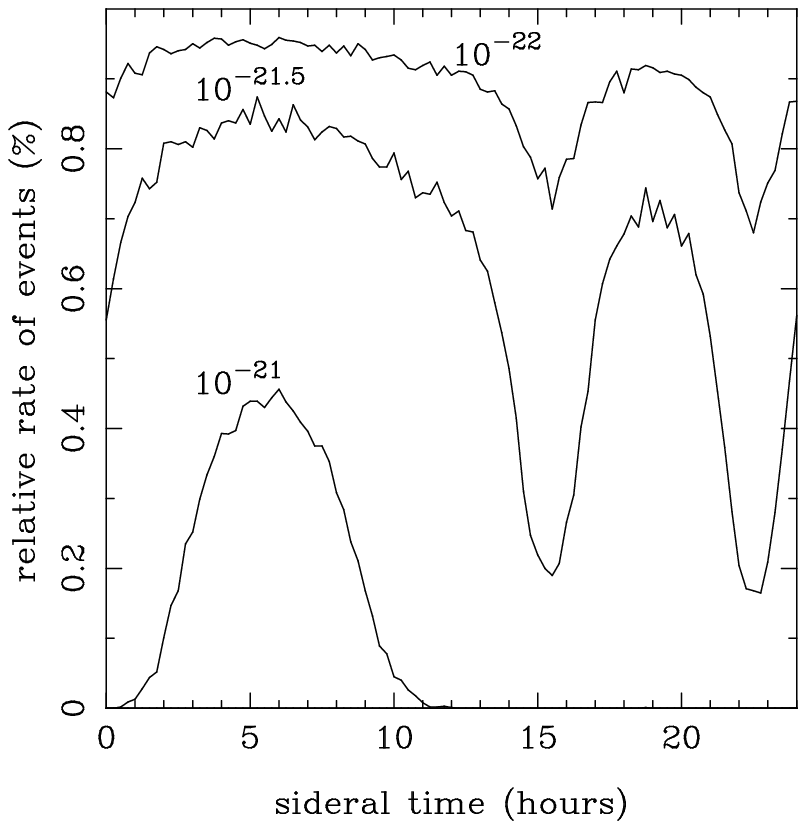}}
\caption{Sources localized in the Galactic center.
See section 3 for explanations.}
\label{fi4}
\end{figure}

\begin{figure}
\resizebox{\hsize}{!}{\includegraphics{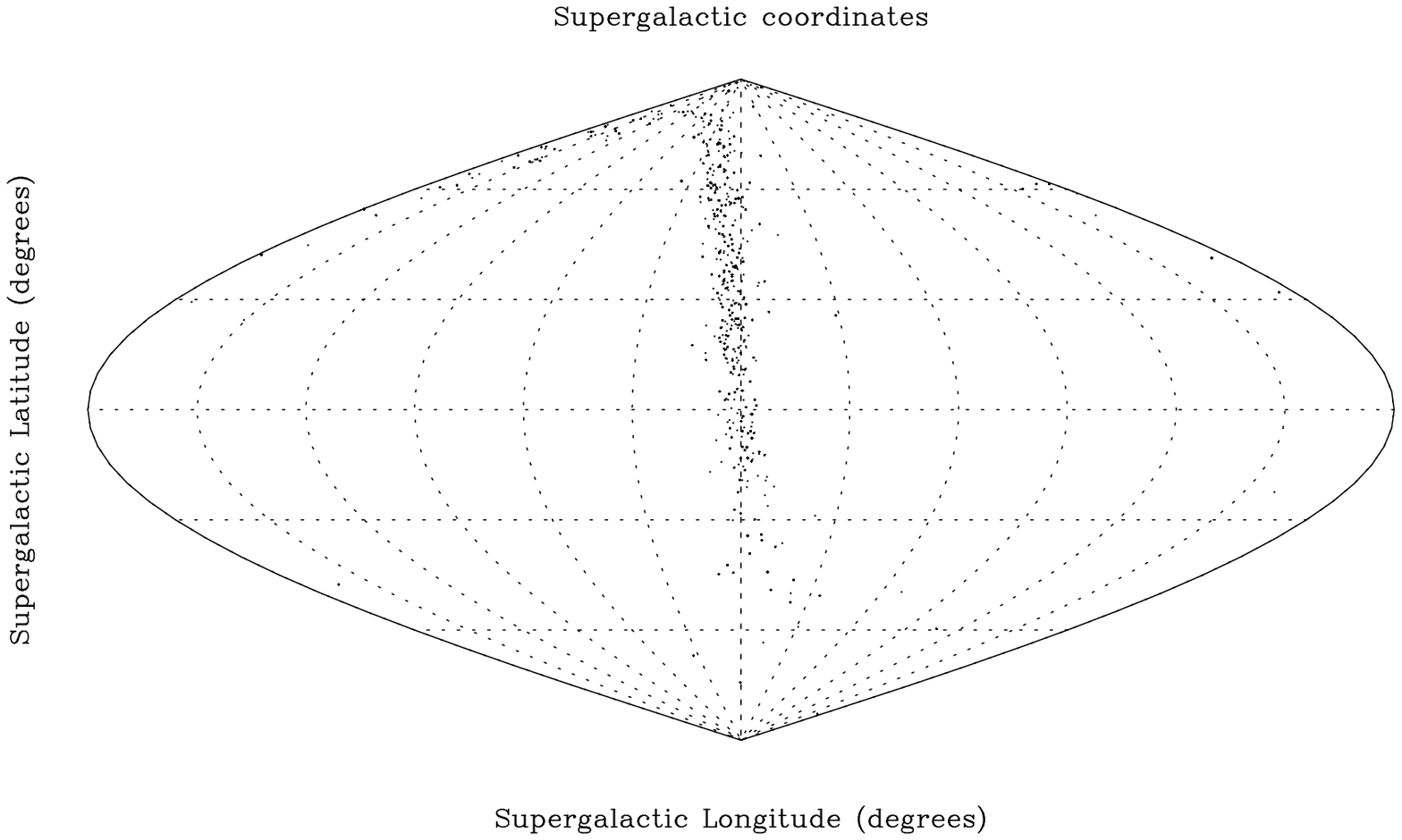}}
\resizebox{\hsize}{!}{\includegraphics{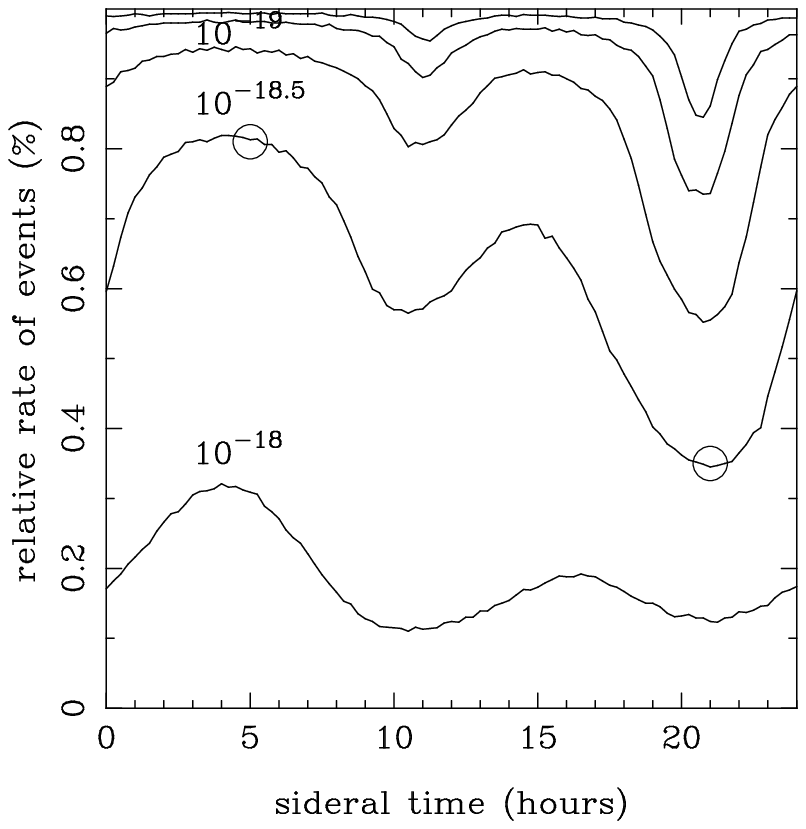}}
\resizebox{\hsize}{!}{\includegraphics{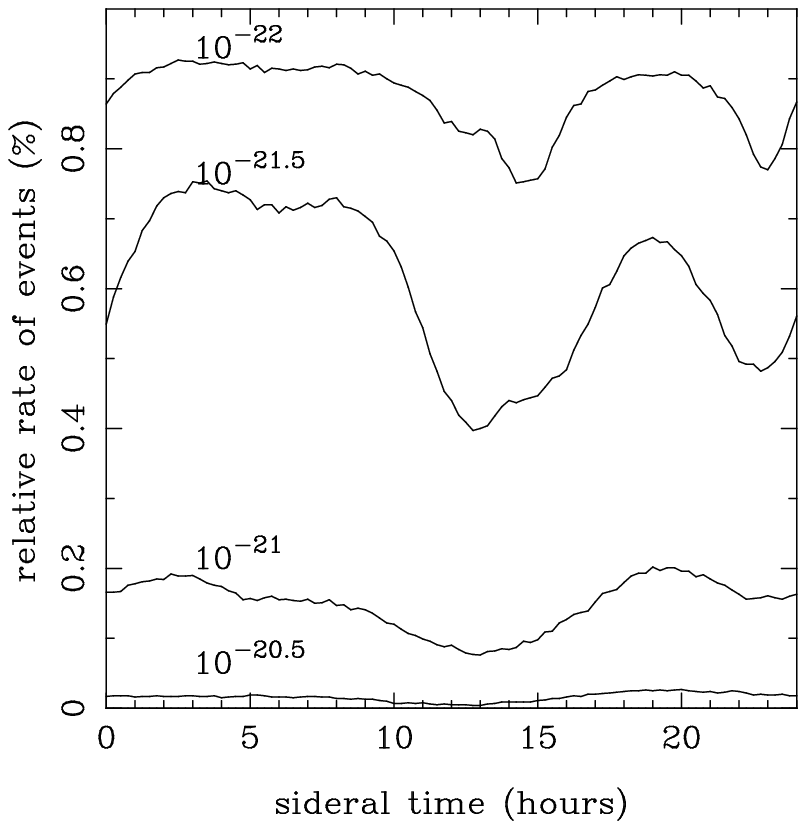}}
\caption{Homogeneous distribution of sources in the galactic plane. 
See section 3 for explanations.}
\label{fi2}
\end{figure}

\begin{figure}
\resizebox{\hsize}{!}{\includegraphics{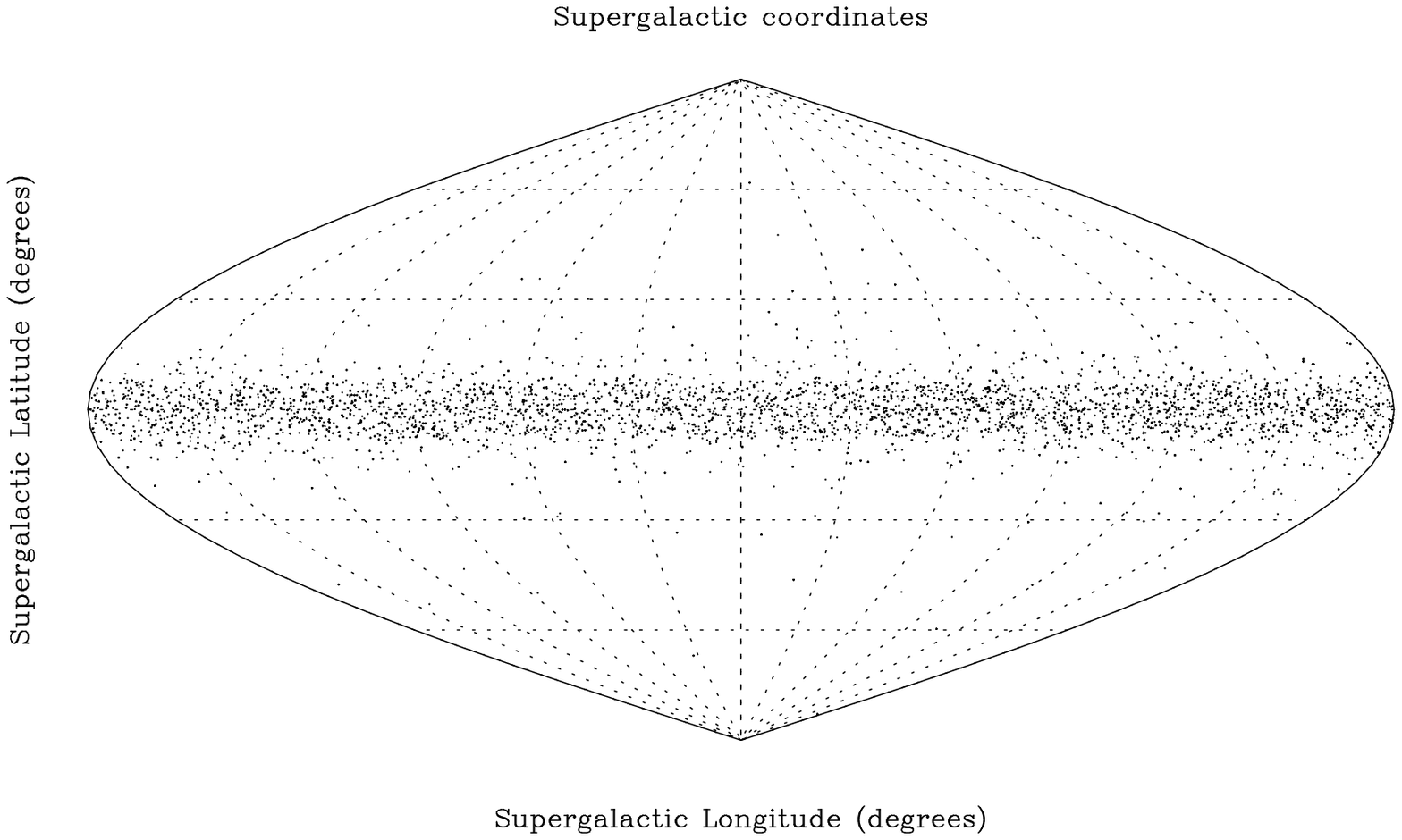}}
\resizebox{\hsize}{!}{\includegraphics{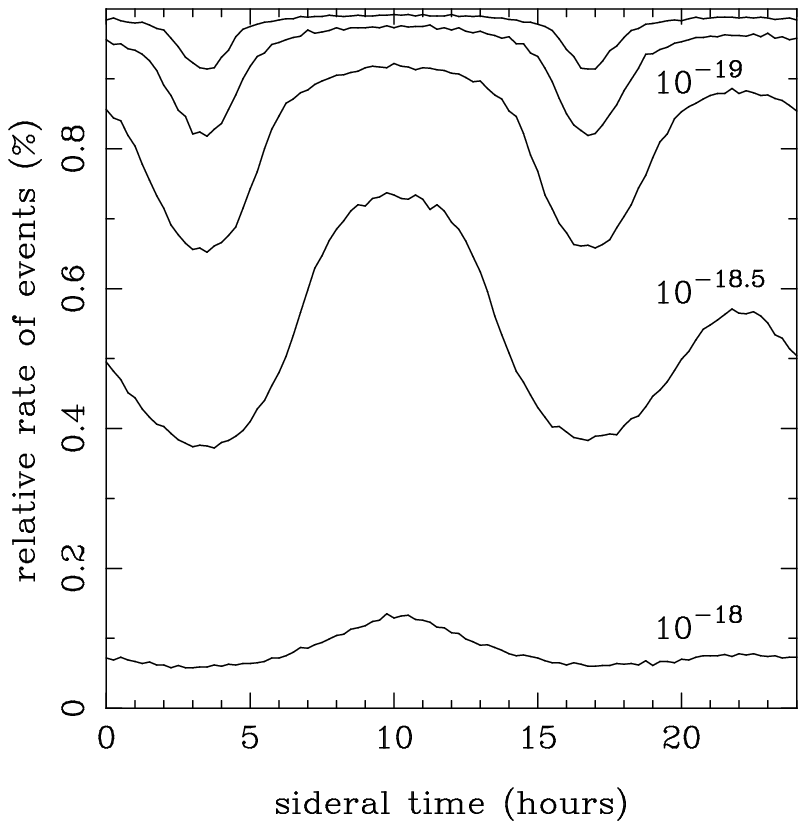}}
\resizebox{\hsize}{!}{\includegraphics{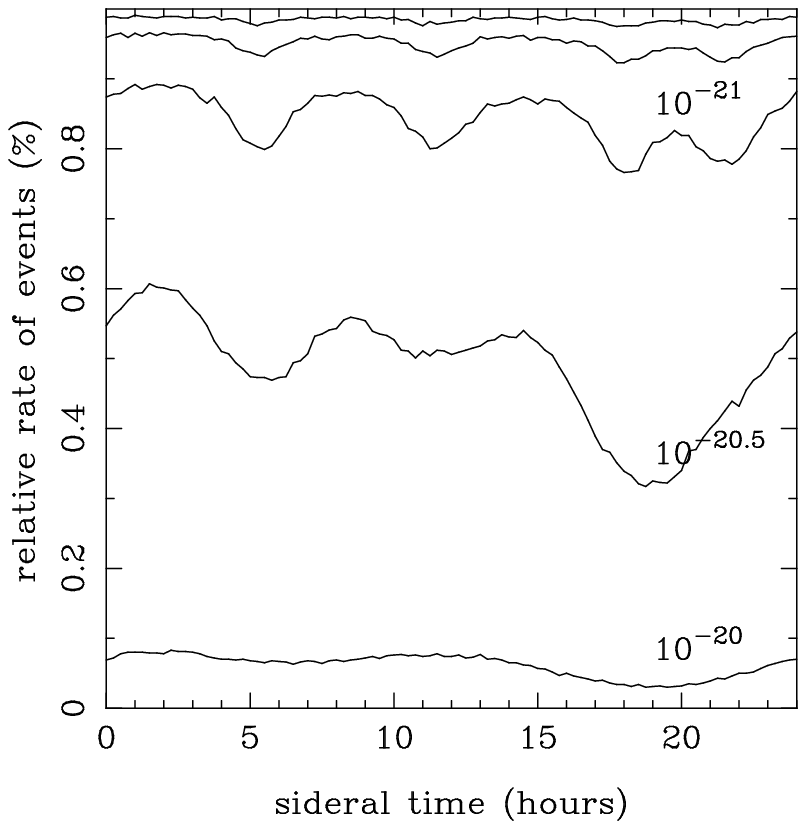}}
\caption{Homogeneous distribution of sources in the supergalactic plane. 
See section 3 for explanations.}
\label{fi3}
\end{figure}

\begin{figure}
\resizebox{\hsize}{!}{\includegraphics{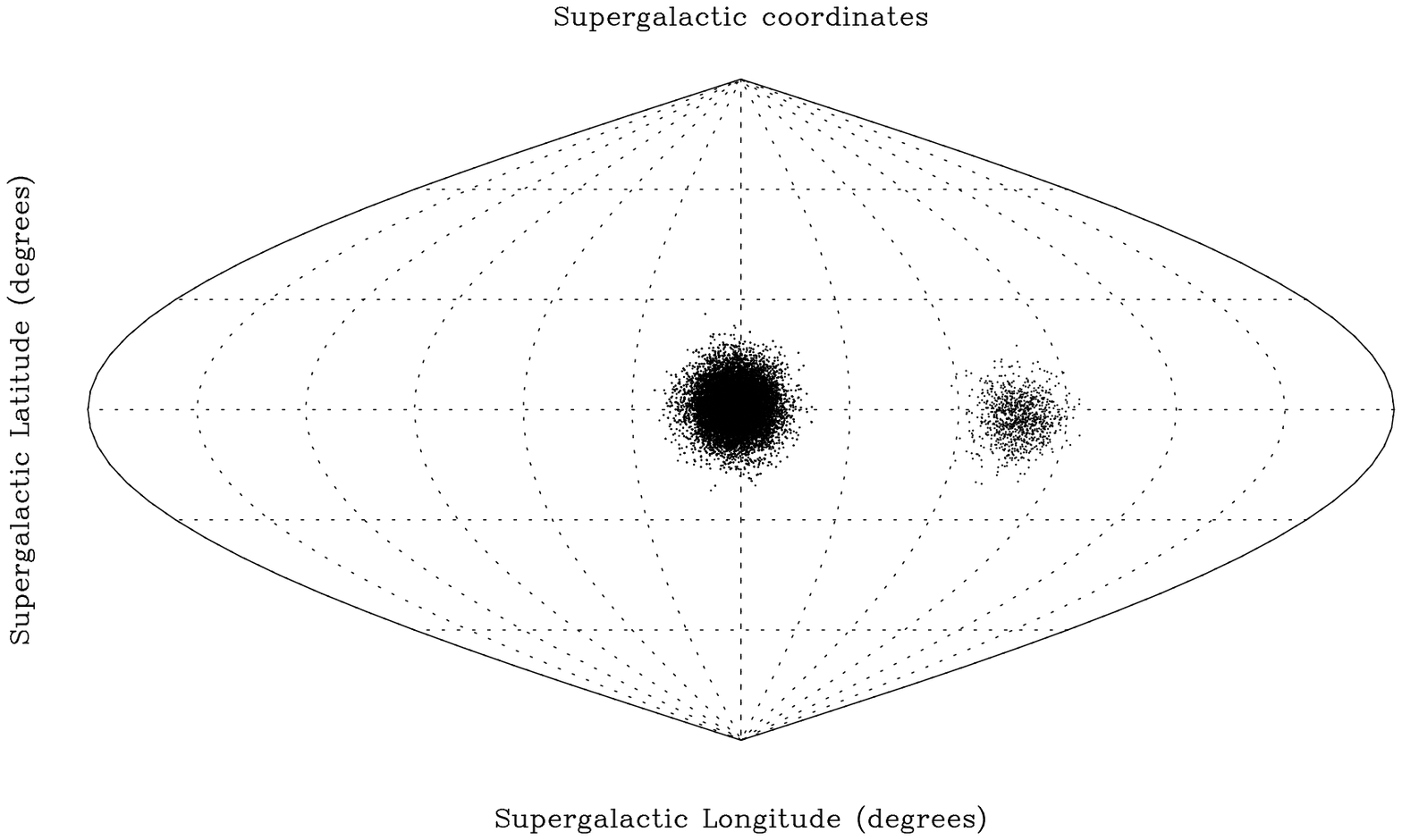}}
\resizebox{\hsize}{!}{\includegraphics{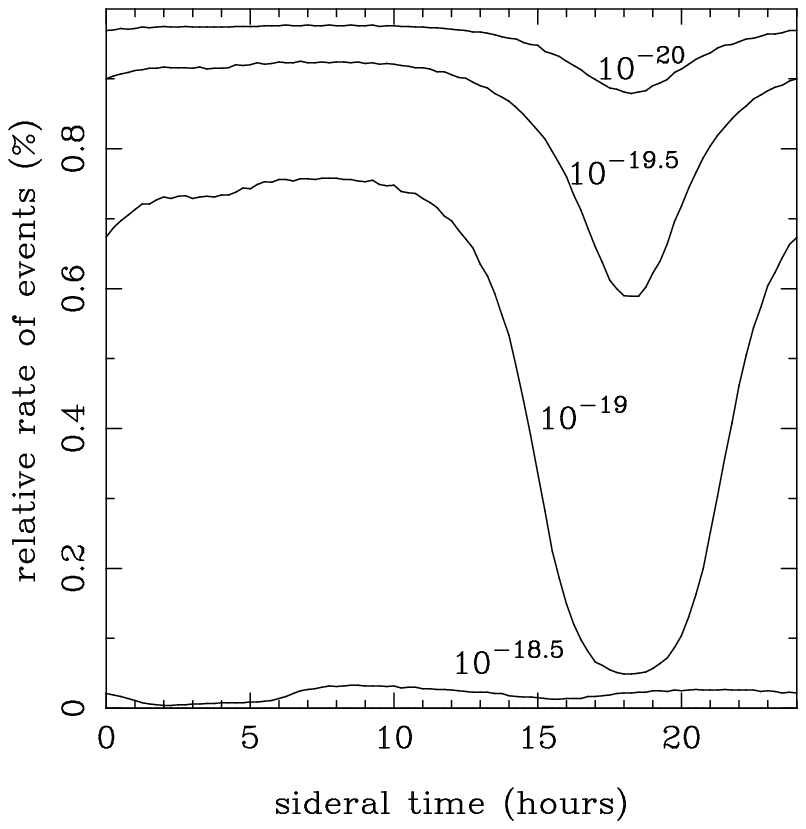}}
\resizebox{\hsize}{!}{\includegraphics{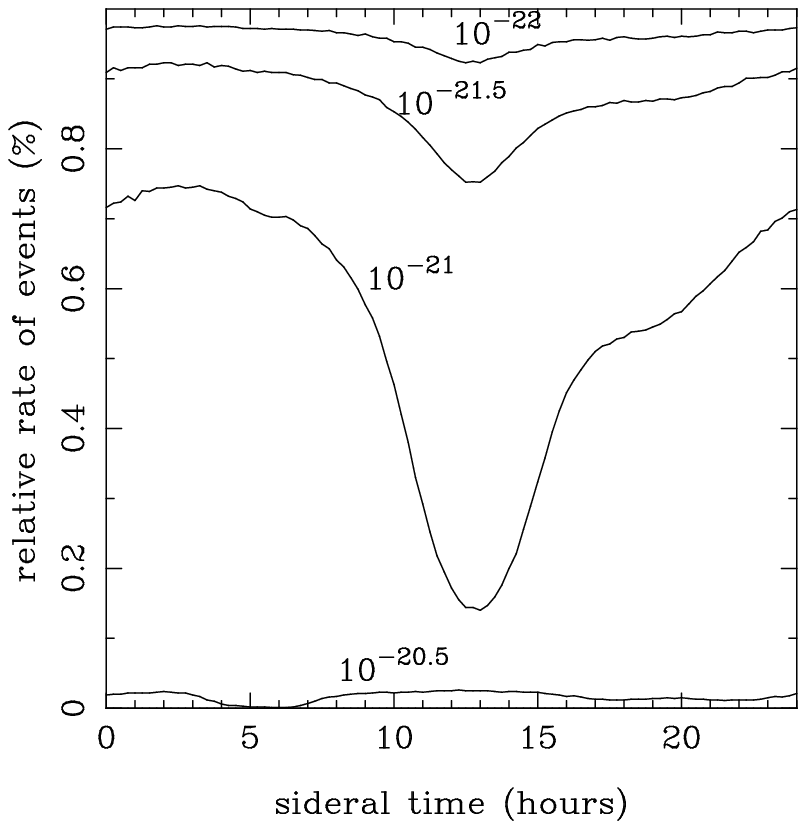}}
\caption{Sources localized in the Virgo cluster and in the Great-Attractor. 
See section 3 for explanations.}
\label{fi5}
\end{figure}

The figures \ref{fi1} to \ref{fi5} give the results for respectively:
\begin{itemize}
\item A 3D homogeneous distribution of sources within 10 kpc, 
\item Sources localized in the Galactic center,
\item Homogeneous distribution in the galactic plane,
\item Homogeneous distribution in the supergalactic plane, 
\item Sources localized in the Virgo cluster and in the Great-Attractor. 
\end{itemize}

Each curve gives
the relative count number of GW events for equivalent sidereal time intervals
centered at different sidereal times. The $Y$-axis is then a relative number of GW events
per unit of time, i.e. a relative rate of events (in percents). 
We give an example.- the middle panel of Fig. \ref{fi2}. It gives
the rate of GW events for a homogeneous distribution of GW emitters in the
galactic plane, as it would be seen from the EXPLORER bar detector. From the intermediate
sensitivity ($10^{-18.5}$) one can see that the rate of events detected around the sidereal time
$ST=5h$ (80\%) is about twice the rate expected around $ST=21h$ (35\%). 
The corresponding points are marked with an open circle in Fig. \ref{fi2}.
By comparing these kinds of curves with observed ones, it should be possible to have a 
signature of the localization of the GW sources. Then, from the sensitivity 
of the detector, one can place some limits on the radiated GW energy.


\section{Discussion and conclusions}
We have presented a method to analyze the output signal produced
by GW detectors operating over a long time. This 'Sidereal 
Time Analysis' method consists of recording all GW events as a function of 
sidereal time, in order to calculate
their rate (number of events per unit of time) as a function of phase, from 0h to 24h,
for different sensitivity thresholds.
If one admits that the noise is uniformly distributed along the sidereal
time, the statistics of observed events can be performed even for 
very low signal to noise ratios (less than one).

A random distribution gives a flat
count rate (Fig. \ref{fi1}). So, any sources distributed in such a 
homogeneous manner and close enough to
be detected will appear as additional noise. Fortunately, the closest sources
would have a galactic origin and would not lead to a flat count rate as
shown by Figs. \ref{fi4} and \ref{fi2}. One can note the similarities between
these two figures. The main characteristic of sources of tensor waves
localized in our Galaxy (Galaxy center or Galaxy disk) is that their count rate
shows a deep depression at 21 h sidereal time for the EXPLORER bar-detector and
at 15h and 22h sidereal time for the VIRGO interferometric detector.

In figure \ref{fi3} we see that the shape of the count rate for sources localized 
in the supergalactic plane shows almost opposite phase in comparison with  Figs. 
\ref{fi4} and \ref{fi2}. This is due to the fact that the supergalactic plane is
almost perpendicular to the galactic plane. This means that one can characterize
the origin of a signal simply from the shape of its count rate. 

Finally, the last calculations for the Virgo cluster and the putative GA
also need comment (Fig.\ref{fi5}). 
We consider first the EXPLORER detector (middle panel).
It shows that, at low sensitivity
(lower curves), one will detect the Virgo cluster first, with a shape similar to
the one found for the supergalactic plane (i.e., one peak at 10h sidereal time and
one peak at 23h sidereal time) and that progressively one will detect a
dominant region (between 2h and 12h with a big depression at 18h) when the GA 
dominates.  So, when
the sensitivity of the detectors are improved, it will be possible
to test the existence of the GA. 

For the VIRGO detector (Fig.\ref{fi5} bottom panel) the  effect is similar. 
We see a depression at 5h  when Virgo dominates, and a depression at 13h when 
the GA dominates. 

The next step consists of using real signals to start such an analysis.
In particular, it will be possible to investigate the important case of
coincident detections from multiple detectors and the case where one has only
some events, the critical question being at what number of events can one
have a significant result about the distribution of GW sources.

\acknowledgements{
We thank the anonymous referee for his/her report and for suggesting 
further extensions of the paper.
}


\end{document}